\documentclass[aps,prl,singlecolumn,12pt,superscriptaddress,nofootinbib,nobibnotes,longbibliography]{revtex4-1}
\usepackage{amssymb}
\usepackage{graphicx}
\usepackage{amsmath}
\usepackage{hyperref}
\usepackage{subfigure}
\usepackage{float}
\usepackage{color}
\usepackage{ulem}
\usepackage[utf8]{inputenc}

\begin{document}

\title{The doomsday of black hole evaporation}
\author{Shao-Jiang Wang}
\email{schwang@itp.ac.cn}
\affiliation{CAS Key Laboratory of Theoretical Physics, Institute of Theoretical Physics, Chinese Academy of Sciences, Beijing 100190, China}

\begin{abstract}
By assuming simultaneously the unitarity of the Hawking evaporation and the universality of Bekenstein entropy bound as well as the validity of cosmic censorship conjecture, we find that the black hole evaporation rate could evolve from the usual inverse square law in black hole mass into a constant evaporation rate near the end of the Hawking evaporation before quantum gravity could come into play, inferring a slightly longer lifetime for lighter black holes.
\end{abstract}
\maketitle

The black hole evaporation~\cite{Hawking:1975vcx} is at the center of playground for pursuing the quantum gravity as it explicitly exhibits an apparent paradox between the initially pure state that collapses into a black hole and the mixed (thermal) state of Hawking radiations after evaporation~\cite{Hawking:1976ra}, which directly contradicts with the unitary expectation from AdS/CFT correspondence~\cite{Maldacena:1997re,Witten:1998qj,Gubser:1998bc} (gauge/gravity duality) and classical geometry/quantum information duality~\cite{Nishioka:2009un} (see, e.g. RT formula~\cite{Ryu:2006bv,Hubeny:2007xt}). Recent progresses~\cite{Penington:2019kki,Almheiri:2019qdq} on locating the minimal quantum extremal surface~\cite{Engelhardt:2014gca} and accounting for new nonperturbative saddle points from spacetime wormhole geometries have reproduced the smoking gun of unitarity, that is, the Page curve~\cite{Page:1993wv}. However, the doomsday of black hole evaporation is still vague due to the absence of quantum theory of gravity.  Nevertheless, it would be appealing to just portray the general feature at the very end of the black hole evaporation from several well-motivated beliefs of principles.

The evaporation time~\cite{Page:1976df} for a black hole of mass $M_\mathrm{BH}$ via Hawking radiations~\cite{Hawking:1975vcx} reads
\begin{align}\label{eq:teva1}
t_\mathrm{eva}=\frac{4}{\alpha^2}\left(\frac{M_\mathrm{BH}}{m_\mathrm{Pl}}\right)^3t_\mathrm{Pl}
\end{align}
with $t_\mathrm{Pl}=\sqrt{G\hbar/c^5}$ and $m_\mathrm{Pl}=\sqrt{\hbar c/G}$, where $\alpha$ is a numerical factor that depends on which particle species can be emitted at a significant rate, and can be taken to be its initial value as most of the evaporation time is spent near the original black hole mass.  $\alpha$ should be larger for smaller $M_\mathrm{BH}$ due to a larger number of particle species to be emitted below the Hawking temperature depending on the particle physics beyond the standard model~\cite{Baker:2022rkn}.

If we assume the black hole evaporation is  unitary as expected from AdS/CFT correspondence so that a distant observer could somehow eventually recover the total information contained in the original black hole, then the total information transfer during the full evaporation time scale $t_\mathrm{eva}$ can be estimated by the Bekenstein-Hawking entropy~\cite{Bekenstein:1972tm,Bekenstein:1973ur} as
\begin{align}
S_\mathrm{BH}=\frac{c^3k_\mathrm{B}}{\hbar}\frac{A_\mathrm{H}}{4G}.
\end{align}
Therefore, the average emission rate of Bekenstein-Hawking entropy could be estimated as
\begin{align}\label{eq:dSdt}
\langle\dot{S}_\mathrm{BH}\rangle\equiv\frac{S_\mathrm{BH}}{t_\mathrm{eva}}=k_\mathrm{B}c^3\frac{\pi\alpha^2}{GM_\mathrm{BH}},
\end{align}
in the context of, for example, a Schwarzschild  black hole with $A_\mathrm{H}=4\pi R_s^2$, $R_s=2GM_\mathrm{BH}/c^2$. It turns out as a nice surprise that the reduced Planck constant $\hbar$ is intimately canceled out in the final result, indicating that the average emission rate of black hole entropy should be ignorant of the quantum mechanical details of black hole evaporation, even though both the Bekenstein-Hawking entropy and Hawking evaporation time heavily rely on the underlying quantum nature of black hole.

Assuming the Bekenstein entropy bound~\cite{Bekenstein:1980jp} (see also~\cite{Casini:2008cr,Blanco:2013joa}), Bekenstein has derived a universal upper bound~\cite{Bekenstein:1981zz} on the average rate of information transfer in bits per second,
\begin{align}\label{eq:dIdt}
\dot{I}_\mathrm{max}=\frac{\pi E}{\hbar \ln 2},
\end{align}
from the generalized second law of thermodynamics and the principle of causality, where $E$ is the message energy that carries information. Let us first look at the average rate of information transfer during the whole evaporation process, in which case the total message energy $E$ should be no larger than the black hole mass $E\leq M_\mathrm{BH}c^2$ that is about to be totally evaporated away. If \eqref{eq:dSdt} is universally bounded by \eqref{eq:dIdt} during the whole evaporation,
\begin{align}\label{eq:Mmin}
\langle\dot{S}_\mathrm{BH}\rangle\leq\dot{I}_\mathrm{max}k_\mathrm{B}\ln 2\Rightarrow M_\mathrm{BH}\geq\alpha m_\mathrm{Pl},
\end{align}
then the black hole cannot be totally evaporated away but left with a remnant mass $\alpha$ in unit of the Planck mass, which directly contradicts with our presumption of total evaporation. On the other hand, if one insists a total evaporation of black hole and no black hole remnant~\cite{Chen:2014jwq} should leave behind, then the Bekenstein bound on the maximum average rate of information transfer cannot be universal but violated at the very end of the black hole evaporation.

To be more rigorous, let us then further look into the average rate of information transfer for a sufficiently short time interval $\Delta t$, during which the black hole mass is reduced from $M_\mathrm{BH}+\Delta M$ into $M_\mathrm{BH}$ by emitting a single Hawking-radiation particle of mass $\Delta M$ and $\alpha(M)$ can be treated as a constant between two mass-step thresholds. It is easy to solve for the evaporated mass as $\Delta M/m_\mathrm{Pl}=[(M_\mathrm{BH}/m_\mathrm{Pl})^3+(\alpha^2/4)(\Delta t/t_\mathrm{Pl})]^{1/3}-(M_\mathrm{BH}/m_\mathrm{Pl})$ from the usual Hawking evaporation rate $\mathrm{d}(M_\mathrm{BH}/m_\mathrm{Pl})/\mathrm{d}(t/t_\mathrm{Pl})=-(\alpha^2/12)(M_\mathrm{BH}/m_\mathrm{Pl})^{-2}$. Then, the instant reduction rate of the black hole entropy by emitting a single Hawking-radiation particle of mass $\Delta M$ can be expanded as
\begin{align}\label{eq:DSDt}
\langle\dot{S}_\mathrm{BH}\rangle\equiv\frac{4\pi k_\mathrm{B}[(M_\mathrm{BH}+\Delta M)^2-M_\mathrm{BH}^2]}{m_\mathrm{Pl}^2\Delta t}=\frac{2\pi\alpha^2k_\mathrm{B}}{3(M_\mathrm{BH}/m_\mathrm{Pl})t_\mathrm{Pl}}+\mathcal{O}(\Delta t).
\end{align}
On the other hand, the upper bound on the average rate of information transfer~\eqref{eq:dIdt} admits a maximal value for this instant emission process since the message energy $E$ from emitting a Hawking-radiation particle cannot be larger than the Planck mass in the semi-classical regime,
\begin{align}\label{eq:DIDt}
\dot{I}_\mathrm{max}k_\mathrm{B}\ln2=\frac{\pi k_\mathrm{B}E}{\hbar}\leq\frac{\pi k_\mathrm{B}m_\mathrm{Pl}c^2}{\hbar}.
\end{align}
Apparently, the instant reduction rate of the black hole entropy~\eqref{eq:DSDt} would eventually diverge as decreasing the black hole mass, and hence there must be a lower bound on the black hole mass after evaporation in order not to violate the upper bound on the information transfer rate~\eqref{eq:DIDt},
\begin{align}
\frac{2\pi\alpha^2k_\mathrm{B}}{3(M_\mathrm{BH}/m_\mathrm{Pl})t_\mathrm{Pl}}\leq\frac{\pi k_\mathrm{B}m_\mathrm{Pl}c^2}{\hbar}\Rightarrow M_\mathrm{BH}\geq\frac23\alpha^2m_\mathrm{Pl}.
\end{align}
Again, if one still insists a full evaporation of black hole leaving no remnant behind, then the maximum bound on the average rate of information transfer cannot be universal but violated near the end of the black hole evaporation.

Nevertheless, if one still wants to preserve the universality of information transfer bound $\langle\dot{S}_\mathrm{BH}\rangle\leq\dot{I}_\mathrm{max}k_\mathrm{B}\ln2$ at the very end of the black hole evaporation, then we can only expect a slower evaporation rate inversely proportional to the black hole mass to a power no larger than 2,
\begin{align}
\frac{\mathrm{d}M}{\mathrm{d}t}=-\lambda\left(\frac{m_\mathrm{Pl}}{t_\mathrm{Pl}}\right)\left(\frac{m_\mathrm{Pl}}{M}\right)^n,\quad 0\leq n<2,
\end{align}
leading to a longer evaporation time near the end of the black hole evaporation,
\begin{align}
t_\mathrm{eva}=\frac{1}{(n+1)\lambda}\left(\frac{M_\mathrm{BH}}{m_\mathrm{Pl}}\right)t_\mathrm{Pl},
\end{align}
during which the universal bound $\langle\dot{S}_\mathrm{BH}\rangle\leq\dot{I}_\mathrm{max}k_\mathrm{B}\ln2$ simply implies an inequality
\begin{align}
M_\mathrm{BH}\geq\left(\frac{(n+1)\lambda c^{4+n/2}\hbar^{n/2}}{G^{2+n/2}}\right)^{\frac{1}{n+2}}\left(\frac{A_\mathrm{H}}{4\pi}\right)^{\frac{1}{n+2}}.
\end{align}
The minimal requirement for this modified period of slower evaporation is to at least preserve the Penrose inequality in order not to expose the singularity. This exactly corresponds to a constant evaporation rate with $n=0$ and $\lambda=1/4$, leading to a longer evaporation time scale of form
\begin{align}\label{eq:teva2}
t_\mathrm{eva}=4\left(\frac{M_\mathrm{BH}}{m_\mathrm{Pl}}\right)t_\mathrm{Pl}
\end{align}
for the black hole mass $M_\mathrm{BH}\leq\alpha m_\mathrm{Pl}$, in which case the universal bound $\langle\dot{S}_\mathrm{BH}\rangle\leq\dot{I}_\mathrm{max}k_\mathrm{B}\ln2$ exactly reproduces the Penrose inequality~\cite{1973NYASA.224..125P,Ben-Dov:2004lmn} (see the review~\cite{Mars:2009cj} for references therein),
\begin{align}
M_\mathrm{BH}\geq\frac{c^2}{G}\sqrt{\frac{A_\mathrm{H}}{16\pi}},
\end{align}
so that the naked singularity would never reveal itself to jeopardize the real world along all the way at least near the final stage of evaporation according to the cosmic censorship conjecture~\cite{Penrose:1969pc,Hawking:1970zqf}. Even without help of cosmic censorship conjecture, the Penrose inequality can be also expected from the AdS/CFT correspondence~\cite{Engelhardt:2019btp,Xiao:2022obq} as a new swampland condition~\cite{Folkestad:2022dse}. Note that the Schwarzschild black hole exactly saturates the Penrose inequality and hence should also saturate the information transfer bound correspondingly.

\begin{figure}[h]
\centering
\includegraphics[width=0.7\textwidth]{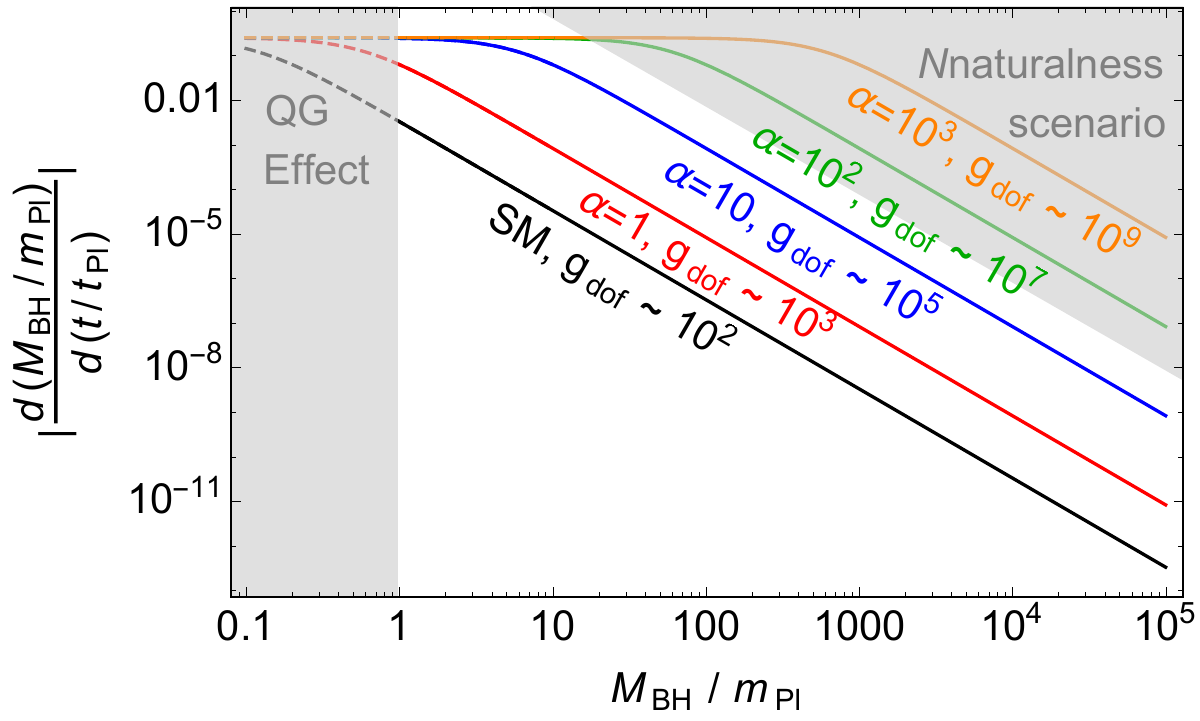}\\
\caption{The absolute evaporation rate for some illustrative values of $\alpha$ with the corresponding number of degrees of freedom estimated at orders of magnitude from $\alpha\sim\mathcal{O}(10^{-3}g_\mathrm{dof})^{1/2}$. The left gray-shaded region surfers from the quantum gravity effect while the upper-right gray-shaded region could be  originated from the $N$naturalness scenario.}\label{fig:EvaporationRate}
\end{figure}

Therefore, the black hole mass should decrease at a rate of form
\begin{align}
\frac{\mathrm{d}M_\mathrm{BH}}{\mathrm{d}t}=-\frac14\left(\frac{m_\mathrm{Pl}}{t_\mathrm{Pl}}\right)\frac{1}{1+3\left(\frac{M_\mathrm{BH}}{\alpha m_\mathrm{Pl}}\right)^2},
\end{align}
which reduces to the usual evaporating rate $M_\mathrm{BH}^{-2}$ for a large black hole mass $M_\mathrm{BH}\gg\alpha m_\mathrm{Pl}$ but starts to approach a constant evaporation rate when $m_\mathrm{Pl}\ll M_\mathrm{BH}<\alpha m_\mathrm{Pl}$. Here we are particularly interested in the case with $M_\mathrm{BH}\gg m_\mathrm{Pl}$ to evade the quantum gravity effect beyond all the above semi-classical treatment. Thus, the total evaporation time scale reads
\begin{align}
t_\mathrm{eva}=4\alpha\left[\left(\frac{M_\mathrm{BH}}{\alpha m_\mathrm{Pl}}\right)+\left(\frac{M_\mathrm{BH}}{\alpha m_\mathrm{Pl}}\right)^3\right]t_\mathrm{Pl},
\end{align}
reproducing \eqref{eq:teva1} and \eqref{eq:teva2} for $M_\mathrm{BH}\gg\alpha m_\mathrm{Pl}$ and $m_\mathrm{Pl}\ll M_\mathrm{BH}\ll\alpha m_\mathrm{Pl}$, respectively. See Fig.~\ref{fig:EvaporationRate} for the absolute evaporation rate for some illustrative values of $\alpha$ with the corresponding number of degrees of freedom estimated at orders of magnitude from $\alpha\sim\mathcal{O}(10^{-3}g_\mathrm{dof})^{1/2}$~\cite{Ukwatta:2015iba}. The left gray-shaded region suffers from the quantum gravity effect while the upper-right gray-shaded region could be originated from the $N$naturalness scenario as discussed below.

Here the requirement $\alpha\gg1$ is not that radical to have a large number of degrees of freedom beyond the standard model (SM) of particle physics as anticipated from~\cite{Baker:2022rkn}, for example, the $N$naturalness scenario~\cite{Arkani-Hamed:2016rle} with upto $N\sim10^{16}$ copies of SM to fully solve the hierarchy problem so that $\alpha\sim10^7$ from $g_\mathrm{dof}\sim Ng_\mathrm{SM}\sim10^{18}$, the scenario of Dvali~\cite{Dvali:2007hz} with $N\sim10^{32}$ copies of SM, and the scenario of large extra dimensions with $10^{32}$ degrees of freedom in the form of Kaluza-Klein gravitons. Nevertheless, if there is no new particle beyond the electroweak scale, then we can compute $\alpha\approx 0.224$ so that we do not need to alter the evaporation rate at all in the first place since the quantum gravity effect would already come into play before the breakdown of the information transfer bound. If not, our final result would necessarily lead to a slightly longer lifetime for lighter black holes. 

To conclude, we have showed in this letter that, when a  vast number of particles is expected beyond the SM, the black hole evaporation rate evolves from the usual inverse square law in mass into a constant rate, provided that the black hole evaporation is unitary and obeys the Bekenstein's upper bound on the average rate of information transfer as well as the Penrose's inequality from the cosmic censorship conjecture. 
This reduced evaporation rate of black hole via emitting Hawking radiations might be caused by some extra energy costs other than the emitted particles to build up wormhole geometries between the island and Hawking radiations in order to recover the unitarity of black hole evaporation.
Alternatively, it is also possible (though less likely) that some of the guiding principles in use might as well be violated near the very end of black hole evaporation even before quantum gravity taking over, providing valuable clues for investigating the late-time breakdown of either quantum mechanics (unitary evaporation), classical information (maximum transmission), or general relativity (hidden singularity).  Either way, we can sketch the unexpected portrait of the doomsday of black hole evaporation.

\begin{acknowledgments}
We thank helpful discussions with Rong-Gen Cai, Song He, Li Li, Shan-Ming Ruan, Jie-Qiang Wu, and Run-Qiu Yang.
This work is supported by the National Key Research and Development Program of China Grants  No. 2021YFC2203004 and No. 2021YFA0718304, 
the National Natural Science Foundation of China Grants No. 12105344 and No. 12235019, 
the Key Research Program of the Chinese Academy of Sciences (CAS) Grant No. XDPB15, 
the Key Research Program of Frontier Sciences of CAS, 
and the Science Research Grants from the China Manned Space Project with No. CMS-CSST-2021-B01.
\end{acknowledgments}

\bibliography{ref}

\end{document}